\documentclass{PoS}
\usepackage{amsmath}
\usepackage{bm}
\graphicspath{{./Figs/}}
\newcommand{\APF}{\mathrm{APF}}
\newcommand{\Zpq}{\mathcal{Z}_\mathrm{pq}}

\title{Evading the model sign problem in the PNJL model with repulsive vector-type interaction via path optimization\footnote{Report No.: YITP-19-109}}

\ShortTitle{Evading the model sign problem in PNJL with vector interaction}

\author{\speaker{Akira Ohnishi}\\ 
	Yukawa Institute for Theoretical Physics, Kyoto University,
	Kyoto 606-8502, Japan\\
        E-mail: \email{ohnishi@yukawa.kyoto-u.ac.jp}}

\author{Yuto Mori\\ 
	Department of Physics, Faculty of Science, Kyoto University,
	Kyoto 606-8502, Japan\\
        E-mail: \email{mori.yuto.47z@st.kyoto-u.ac.jp}}

\author{Kouji Kashiwa\\ 
	Fukuoka Institute of Technology, Wajiro, Fukuoka 811-0295, Japan\\
	Department of Physics, Faculty of Science, Kyoto University,
	Kyoto 606-8502, Japan\\
        E-mail: \email{kashiwa@fit.ac.jp}}

\abstract{
We discuss the sign problem in the Polyakov loop extended Nambu--Jona-Lasinio model
with repulsive vector-type interaction by using the path optimization method.
In this model, both of the Polyakov loop and the vector-type interaction
cause the model sign problem, and several prescriptions have been utilized
even in the mean field treatment.
In the path optimization method, integration variables are complexified
and the integration path (manifold) is optimized to evade the sign problem,
or equivalently to enhance the average phase factor.
Within the homogeneous field ansatz, the path is optimized by using the feedforward neural network.
We find that the assumptions adopted in previous works,
$\mathrm{Re}\,A_8 \simeq 0$ and $\mathrm{Re}\,\omega \simeq 0$,
can be justified from the Monte-Carlo configurations sampled on the optimized path.
We also derive the Euler-Lagrange equation for the optimal path to satisfy.
The two optimized paths, 
the solution of the Euler-Lagrange equation and the variationally optimized path,
agree with each other in the region with large statistical weight.
}

\FullConference{37th International Symposium on Lattice Field Theory - Lattice2019\\
		16-22 June 2019\\
		Wuhan, China}

\begin{document}

\section{Introduction}

Elucidating the structure of the QCD phase diagram is
one of the important challenges in high-energy and nuclear physics.
For example, the first order QCD phase transition boundary may exist
at finite baryon density,
and, if exists, it will affect the dynamics of dense matter
realized during heavy-ion collisions and compact astrophysical phenomena.
In fact,
recent observation of the negative slope in the directed flow, $dv_a/dy<0$,
at colliding energies at
$\sqrt{s_{{\scriptscriptstyle NN}}}\simeq 10~\mathrm{GeV}$~\cite{STARv1}
suggests the softening of the equation of state at high densities~\cite{Nara}.
By comparison, the first order phase boundaries
do exist in the imaginary chemical potentials
$\theta\equiv \mathrm{Im}\mu_q/T=\pi/3, \pi, 5\pi/3, \ldots$
at high temperatures~\cite{RW}.
This phase transition appears from the Roberge-Weiss periodicity,
has the $Z_3$ origin, and is relevant to the confinement-deconfinement
transition.
If these boundaries are connected in the complexified $\mu_B$ space,
therefore, finite baryon density (finite $\mathrm{Re}\mu_B$) phase transition
can be interpreted
as the deconfinement assisted chiral phase transition~\cite{Kashiwa2015}.

In order to discuss these aspects of the QCD phase diagram,
we need calculations at finite real part of chemical potential,
where the sign problem exists in the lattice QCD simulation
and even in the mean field treatment of chiral effective models
with the deconfinement effects,
such as the Polyakov loop extended Nambu--Jona-Lasino (PNJL) model~\cite{PNJL}.
%
Let us mention here the model sign problem
(the sign problem appearing in the QCD effective models)
in the PNJL model.
First, the Polyakov loop causes the model sign problem.
In the diagonalized gauge, the Polyakov loop (its conjugate) is given as
$\Phi=\mathrm{Tr}(U)/N_c$ ($\overline{\Phi}=\mathrm{Tr}(U^{-1}))/N_c$)
where $U=\exp(i(A_3\lambda_3+A_8\lambda_8)/T)$ is the temporal link variable
and $A_3$ and $A_8$ are the color components of the temporal gluon field.
At finite chemical potential, the Boltzmann weight becomes complex
and $\Phi$ and $\overline{\Phi}$ have different expectation values.
In order to simulate the difference of $\Phi$ and $\overline{\Phi}$
in the mean field approximation,
one usually imposes stationary conditions for the effective potential
with respect to $\Phi$ and $\overline{\Phi}$
as if $\Phi$ and $\overline{\Phi}$ take independent values.
As long as the weight is real and $A_3$ and $A_8$ take real values,
however, $\Phi$ and $\overline{\Phi}$ are not independent
but take complex conjugate values.
One of the promising ideas to justify the mean field treatment
is to assume that the $\mathcal{CK}$ symmetry exists
in the fermion determinant at finite density~\cite{CKsym},
where $\mathcal{C}$ and $\mathcal{K}$ are
the charge conjugation and the complex conjugation, respectively.
In the $\mathcal{CK}$ symmetry ansatz, one assumes 
real $A_3$ and pure imaginary $A_8$ values 
($\mathrm{Im}A_3=0$ and $\mathrm{Re}A_8=0$).
These constraints justify the above mentioned mean field treatment.
%
Second, the repulsive vector-type interaction also induces the model sign problem.
In the Hubbard-Stratonovich transformation of the repulsive interaction,
one generally get a complex effective action term,
$\exp[-(\bar{\psi}\psi)^2]=\int d\omega \exp[-\omega^2+2i\omega(\bar{\psi}\psi)]$.
In the mean-field approximation, the stationary condition often leads
to a pure imaginary value of the auxiliary field $\omega$,
and the effective potential is kept to be real.
However, the stationary point corresponds to the maximum of the thermodynamic potential
along the imaginary direction of the auxiliary field and is not stable, in principle.
Thus it would be desirable to examine if the above mean field ansatz is realized.

In this proceedings, we discuss the sign problem in the PNJL model
with repulsive vector-type interaction by using the path optimization method~\cite{Kashiwa2019b}.
The path optimization method is one of the complexified variable methods for the sign problem.
The integration path (manifold) is parameterized and optimized variationally to evade the sign problem.
Thus it is possible to examine the assumptions invoked in the mean field treatment,
$\mathrm{Re}\,A_8 \simeq 0$
and 
$\mathrm{Re}\,\omega \simeq 0$
in the Monte-Carlo configurations on the optimized path.

%

\section{Path Optimization Method}
\label{Sec:POM}

There are many approaches to the sign problem as discussed
in the previous and present lattice meetings.
We here concentrate on the complexified variable methods,
the complex Langevin method (CLM)~\cite{CLM,Nagata},
the Lefschetz thimble method (LTM)~\cite{LTM,Mori2018b,Fukuma},
and the path optimization method~\cite{%
Kashiwa2019b,
POM,
Kashiwa2019a}.
%
In CLM, we can sample configurations 
by solving the complex Langevin equation.
Since observables are obtained as the configuration average,
there is no sign problem in successful cases.
However, it should be noted that CLM is not guaranteed to give correct results,
when the distribution of the drift term (derivative of the action)
does not fall off exponentially or faster at large values~\cite{Nagata}.
Since there are many zeros of the Fermion determinant
in the Nambu--Jona-Lasinio model
in the complexified auxiliary fields~\cite{Mori2018b}, 
the drift term distribution would have a power-law tail.
%
In LTM, by solving the holomorphic flow equations,
one can obtain the integral path (manifold) referred to as a thimble
on which the imaginary part of the action is constant.
LTM has been successfully applied to some of field theories,
but it still has several problems.
5Still, LTM has several problems.
One of them comes from the complex phase of the measure (Jacobian),
and referred to as the residual sign problem.
In addition, several thimbles having different imaginary parts
can contribute to the partition function
and weight cancellation can take place.
This problem is called the global sign problem,
which needs special care~\cite{Fukuma}.

The present authors proposed another complexified variable method,
the path optimization method,
in which the parameterized integration path is optimized variationally
to weaken the sign problem
or equivalently to enhance the average phase factor (APF)~\cite{%
Kashiwa2019b,
POM,
Kashiwa2019a}, 
\begin{align}
\APF=\langle e^{i\theta}\rangle_\mathrm{pq}
=\frac{\int_\mathcal{C} d^Nx\, J(z) \exp[-S(z)]}
   {\int_\mathcal{C} d^Nx\, |J(z) \exp[-S(z)]|}
=\frac{\mathcal{Z}}{\Zpq}
\ ,
\end{align}
where $N$ denotes the number of variables
and $J(z)=\det(\partial z_i/\partial x_i)$ is the Jacobian
with $z_i=x_i+iy_i$ being the complexified variable
and $y_i=y_i(x)$ being the imaginary part parameterized as functions
of the real variables $x$.
Provided that the action is a holomorphic (complex analytic) function
of the complexified variable $z$ and the integration path is obtained
by the continuous deformation from the real axis,
the partition function $\mathcal{Z}$ is invariant
while the phase quenched partition function $\Zpq$
depends on the integration path.
Thus enhancing APF corresponds to minimizing $\Zpq$.
In our previous works,
we have parameterized the integration path by some function
or by using the neural network~\cite{POM},
and have optimized the path variationally.
By comparison, since the optimization requires minimizing $\Zpq$
as a functional of $y(x)$,
it is also possible to obtain the optimized path
by solving the Euler-Lagrange equation, $\delta\mathcal{Z}_\mathrm{pq}=0$,
\begin{align}
&\left[
\frac{\partial}{\partial x_i}\,\frac{\partial}{\partial(\partial_iy_j)}
- \frac{\partial}{\partial y_j}
\right]
\left|W(x+iy,\partial y)\right|=0
\quad\mathrm{or}\quad W=0\ ,
\\
&W(x+iy,\partial y)=J(x+iy)\exp\left[-S(x+iy)\right]
=\det(\delta_{ij}+i\partial y_i/\partial x_j)\,\exp[-S(x+iy)]
\ .
\end{align}
In the one variable case,
the Euler-Lagrange equation reads,
\begin{align}
\ddot{y}=(1+{\dot{y}}^2)^2 \left[
\frac{\partial (\mathrm{Im}S)}{\partial x}
+\frac{\dot{y}}{1+{\dot{y}}^2}\frac{\partial (\mathrm{Re}S)}{\partial x}
\right]
\ ,
\end{align}
where $\dot{y}=dy/dx$ and $\ddot{y}=d^2y/dx^2$.

\begin{figure}[htbp]
\begin{center}
\includegraphics[width=7cm]{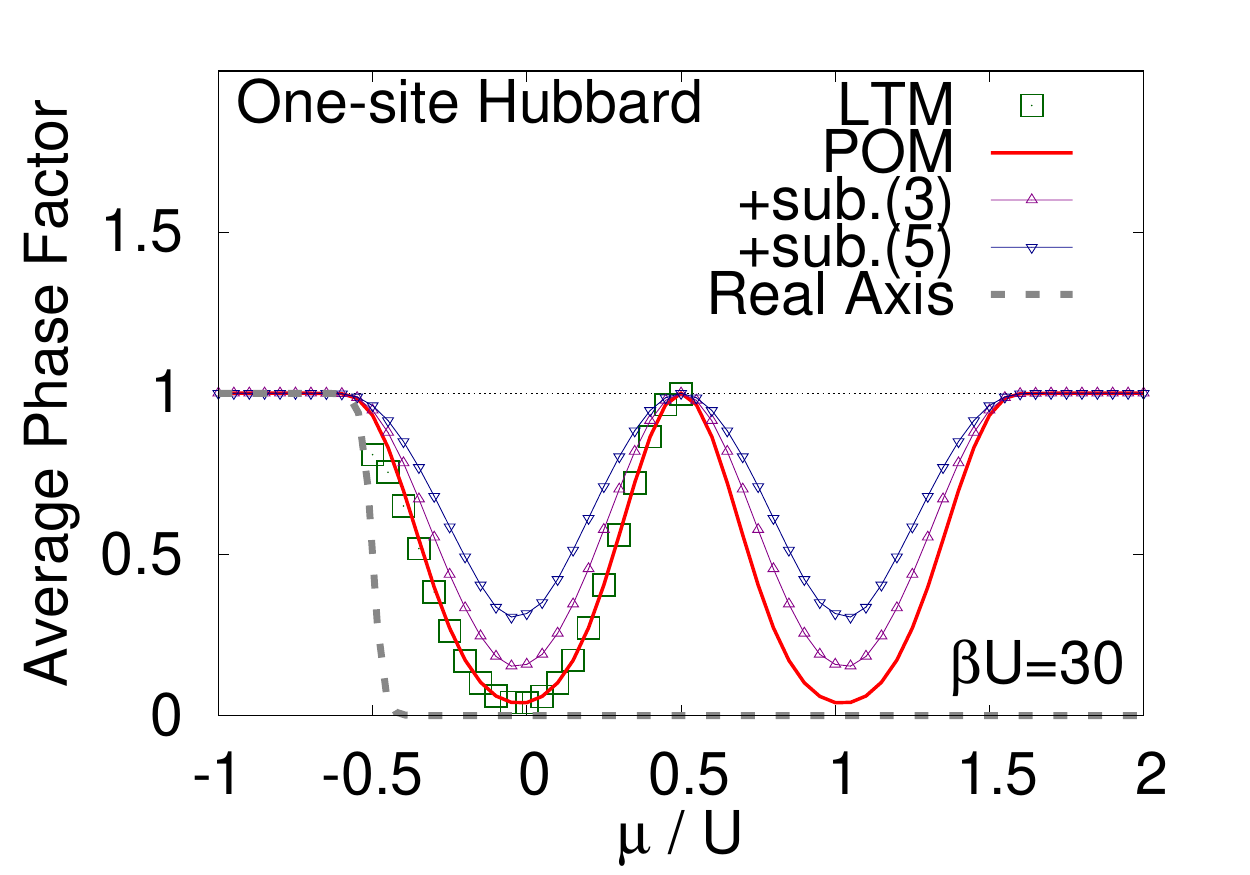}%
\includegraphics[width=7cm]{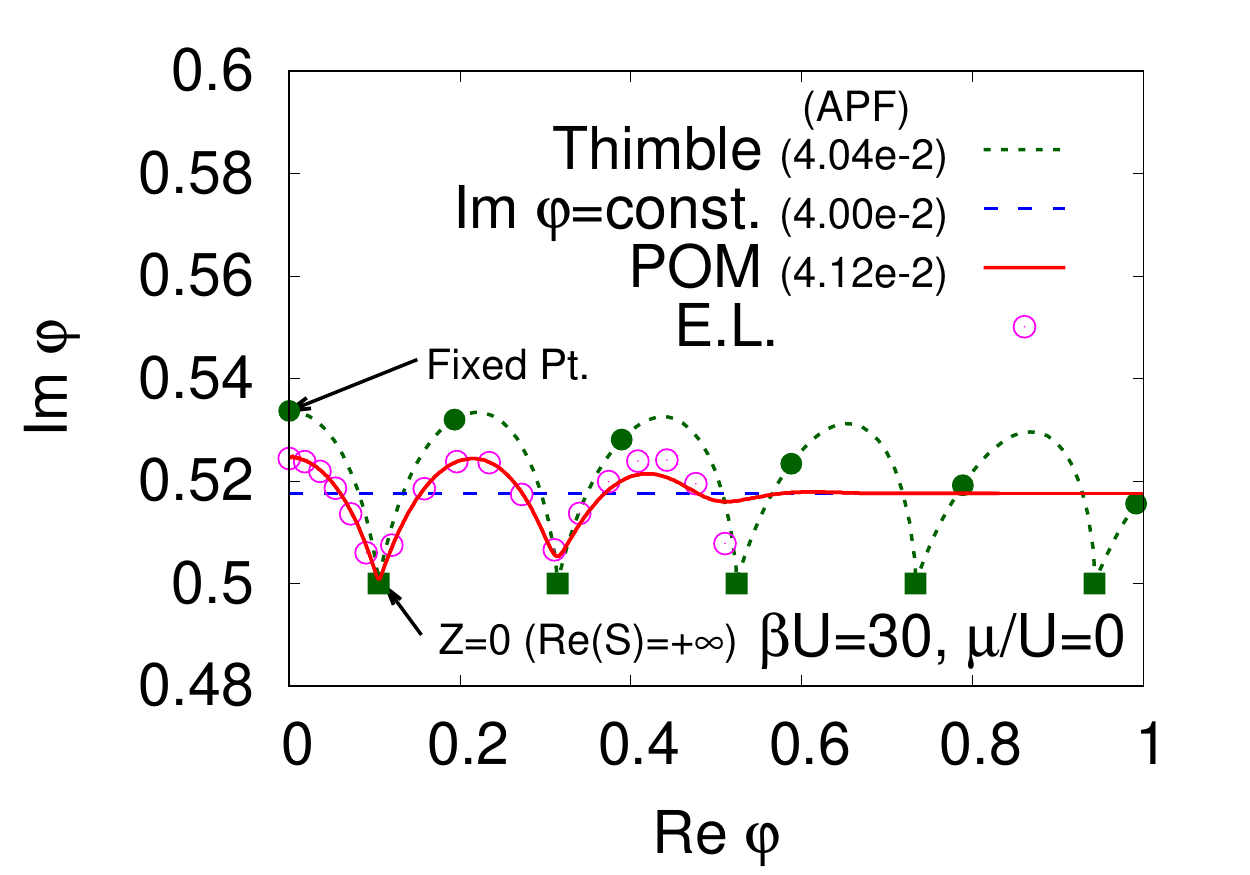}%
\end{center}
\caption{Average phase factor as a function of $\mu/U$ (left)
and the integration path at $\mu/U=0$ (right) in the one-site Hubbard model.}
\label{Fig:Hubbard}
\end{figure}

Let us examine the solution of the Euler-Lagrange equation
in the one-site Hubbard model.
The action and the path integral representation of the partition function
are given as~\cite{Tanizaki},
\begin{align}
S=& U n_\uparrow n_\downarrow - \mu (n_\uparrow+n_\downarrow)
\ ,\\
\mathcal{Z}=&\sqrt{\frac{\beta U}{2\pi}}\int d\varphi 
\left\{
1+\exp[\beta U(i\varphi+\mu/U+1/2)]
\right\}^2\,\exp(-\beta U\varphi^2/2)
\ .
\end{align}
The repulsive interaction between the spin up and down fermions
causes a complex coupling in the Hubbard-Stratonovich transformation.
It is known that this model has a serous sign problem on the real axis
off the half-filling ($\mu/U=0.5$)
as shown in the grey dashed line in the left panel of Fig.~\ref{Fig:Hubbard},
and that the mean field approximation does not work.
In addition, the number of thimbles contributing to the partition function
with different signs increases at lower temperatures~\cite{Tanizaki}.

In the right panel of Fig.~\ref{Fig:Hubbard},
we show
the thimbles (dashed curve),
the variationally optimized path (solid curve),
and the solution of the Euler-Lagrange equation (open circles).
Thimbles are obtained by solving the flow equation,
$d\varphi/dt=\overline{(\partial S/\partial \varphi)}$,
from the fixed points, $\partial S/\partial \varphi=0$ (filled circles).
As for the variationally optimized path,
we start from the real axis,
first optimize the path by assuming constant imaginary part,
and next we optimize the path by using the gradient descent method
to reduce $\Zpq$.
Finding the solution of the Euler-Lagrange equation is more tedious.
We first search for the $\mathrm{Im}\,\varphi$ value at $\mathrm{Re}\,\varphi=0$
so that the solution does not diverge.
The solution
is found to reach the zero point of the statistical weight,
$\varphi_n^{(W=0)}=(2n+1)\pi/\beta{U}+i(\mu/U+1/2)$ with $n=0$.
At the point $W=0$, the derivative can have a gap.
Thus we search for the slope $d(\mathrm{Im}\varphi)/d(\mathrm{Re}\varphi)$
at $\varphi=\varphi_0^{(W=0)}$ such that the solution does not diverge.
So obtained continuous solution of the Euler-Lagrange equation
connects the zero points of the statistical weight, $W=0$,
which are the singular points of the action.
The variationally optimized path is found to agree
with the solution of the Euler-Lagrange equation
in the small $\mathrm{Re}\,\varphi$ region,
where the phase quenched statistical weight $|W|$ is large.
In the large $\mathrm{Re}\,\varphi$ region,
$|W|$ is small and the APF is not sensitive to $\mathrm{Im}\,\varphi$.
It is interesting to find that these optimized paths
do not go through the fixed points of the action.
In POM, the complex phase of the Jacobian cancels
the phase of the Boltzmann weight,
and thus $\mathrm{Im}\,S$ is not necessarily constant along the path.
With this cancellation,
APF on the optimized path is larger than that in LTM,
while the difference is not significant.

It seems that APF becomes very small,
$\APF\simeq 4.12\times 10^{-2}$ at $\mu=0$,
even 
on the optimized path.
This is because the contribution from different thimbles cancel
with each other; even thimbles
($-(2n-1)\pi/\beta{U} \leq \mathrm{Re}\,\varphi \leq (2n+1)\pi/\beta{U}$, 
$n=$ even integer)
give positive integrals,
while the odd thimbles ($n=$ odd integer)
gives negative integrals.
Thus we have a global sign problem in the one-site Hubbard model,
and APF seems to have the upper bound.
This kind of multimodal problem can be weakened
by introducing tempering~\cite{Fukuma}.
After the Lattice 2019 meeting,
we have found that the subset resummation~\cite{Bloch2013} would be also useful 
to go beyond the upper bound.
The partition function can be rewritten as the sum of integrals
on shifted paths,
\begin{align}
\mathcal{Z}=&
\int_\mathcal{C} dx\, W_\mathrm{subset}(x)\ ,\quad
W_\mathrm{subset}= \sum_k w_k J(z+\Delta_k) \exp[-S(z+\Delta_k)]
\ ,\quad \sum_k w_k=1
\ .
\end{align}
The resummed APF,
$\APF_\mathrm{subset}
\equiv \int dx\, W_\mathrm{subset}(x) / \int dx\, |W_\mathrm{subset}(x)|$,
can be larger, if strong cancellation is already taken into account
in the subset.
%
In the one-site Hubbard model,
we define the subset as $k=-1,0,1$ with $\Delta_k=2k\pi/\beta{U}$
and $w_k=1/4$ for $k=\pm1$
to take account of the cancellation in the subset.
Then the resummed APF
becomes larger as shown by triangles in Fig.~\ref{Fig:Hubbard}.
It further increases with resummation of five paths
as shown by inverted triangles.

\section{Application to Polyakov-loop extended NJL (PNJL) model with repulsive vector-type interaction at real $\mu$}
\label{Sec:PNJL}

Now let us proceed to discuss the model sign problem 
appearing in the PNJL model with repulsive vector-type interaction.
The Lagrangian density in the Euclidean spacetime is given as
\begin{align}
\mathcal{L}_E
=&\bar{q}\left[\not\!\!D(\Phi,\overline{\Phi})+m_0\right]q 
- G\left[ (\bar{q}q)^2 + (\bar{q}i\gamma_5 \bm{\tau}q)^2 \right]
+ G_v (\bar{q}\gamma_\mu q)^2
+ \mathcal{V}_g(\Phi,\overline{\Phi})
\ ,
\end{align}
where $m_0$ is the current quark mass,
$D_\nu=\partial_\nu-igA_\nu\delta_{\nu 4}$ is the covariant derivative,
$\Phi (\overline{\Phi})$ is the Polyakov loop (its conjugate),
and $\mathcal{V}_g$ is the gluonic contribution.
By the Hubbard-Stratonovich transformation,
the effective Lagrangian is obtained in the bilinear form of quarks,
\begin{align}
\mathcal{L}_\mathrm{eff}
=&\bar{q}\left[\not\!\!D +m_0
-2G(\sigma+i\gamma_5\bm{\pi}\cdot\bm{\tau})
+2iG_v\gamma_4\omega
\right]q 
+ \mathcal{V}_g(\Phi,\overline{\Phi})
+ G(\sigma^2 + \bm{\pi}^2)
+ G_v \omega^2
\ .\label{Eq:Leff}
\end{align}
We have introduced the scalar ($\sigma$), pseudoscalar ($\bm{\pi}$)
and the temporal component of the vector ($\omega$) auxiliary fields.
It should be noted that the repulsive vector-type interaction induces the model sign problem,
as is clear from the $2iG_v\gamma_4\omega$ term in Eq.~\eqref{Eq:Leff}.
We employ the homogeneous auxiliary-field ansatz as adopted in previous works
using the Monte-Carlo PNJL model~\cite{Kashiwa2019a,Cristoforetti},
which corresponds to the momentum truncation to k = 0,
$\mathcal{Z} =\int \prod_{\bm{k}} dz_{\bm{k}} \exp[-\Gamma(z)]
\simeq \mathcal{N} \int dz_{0} \exp\left[-\Gamma(z_0)\right]$
with 
$\Gamma=\beta V\mathcal{V}_\mathrm{eff}=k\mathcal{V}_\mathrm{eff}/T^4$.
Thus our numerical results converge to the mean-field results in the infinite volume limit.
An explicit form of the effective potential $\mathcal{V}_\mathrm{eff}$ is given in Ref.~\cite{Kashiwa2019b}.


\begin{figure}[htbp]
\begin{center}
\includegraphics[width=5cm,bb=40 25 348 330,clip]{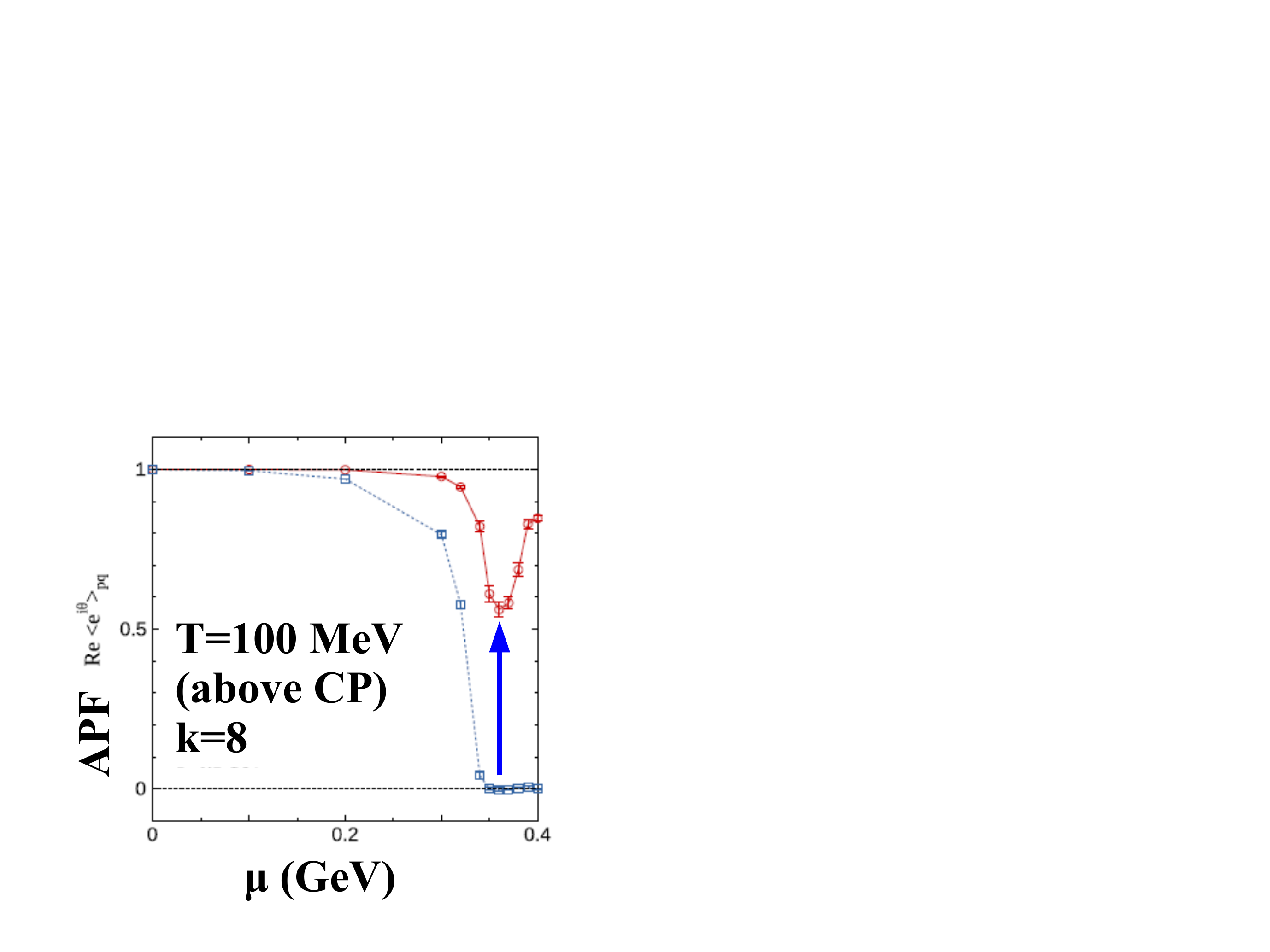}%
\includegraphics[width=10cm,bb=50 33 540 320,clip]{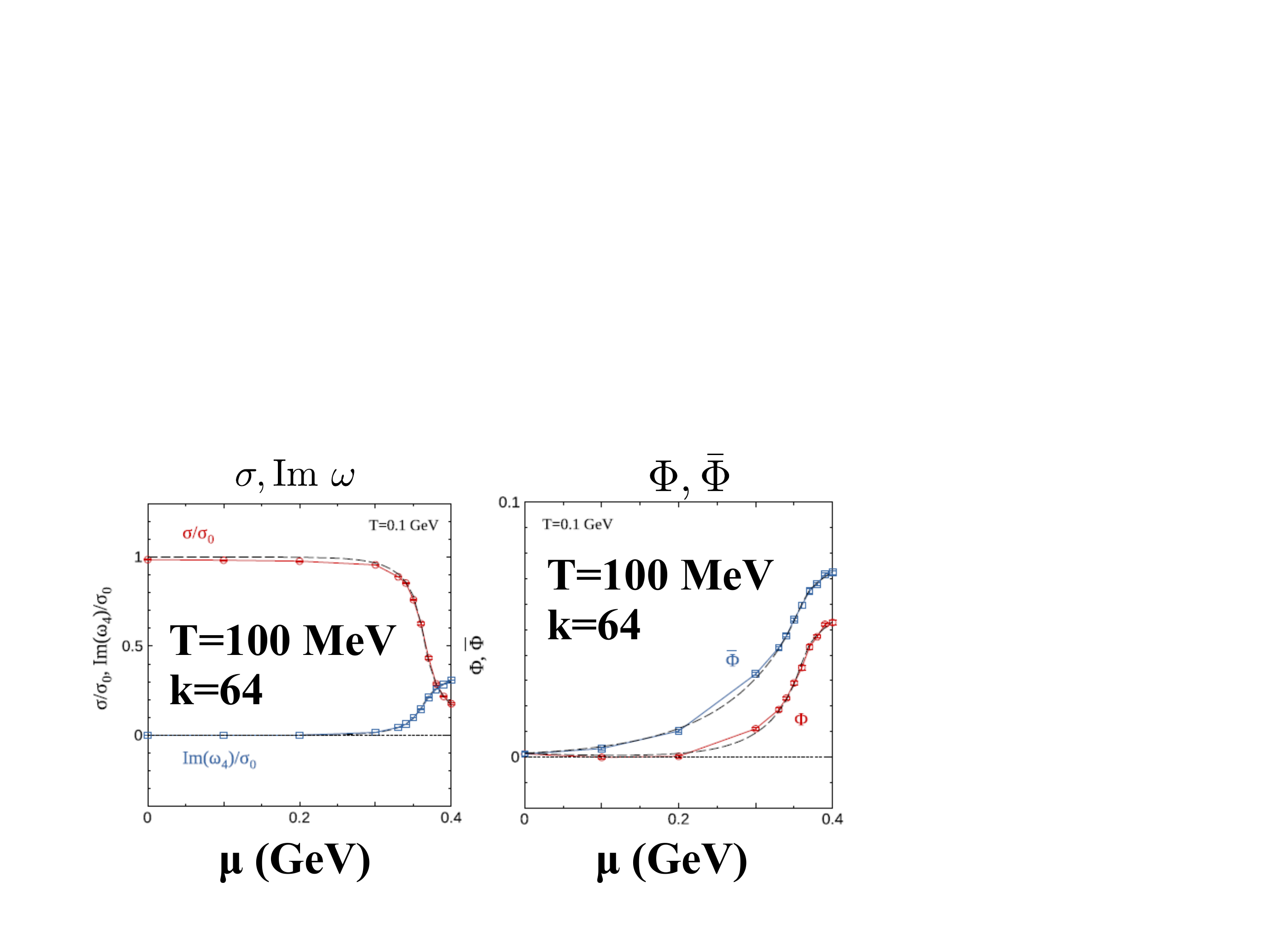}%
\end{center}
\caption{Average phase factor (left), $\sigma$ and $\mathrm{Im}\omega$ (middle)
and $\Phi (\overline{\Phi})$ (right) as functions of quark chemical potential
$\mu$ in the PNJL model.}
\label{Fig:PNJL1}
\end{figure}

We have complexified $A_3$, $A_8$ and $\omega$,
and searched for the integration path by using the feedforward neural network~\cite{Kashiwa2019b}.
In the left panel of Fig.~\ref{Fig:PNJL1}, 
we show APF as a function of the quark chemical potential $\mu$
with (solid) and without (dashed) path optimization.
Without optimization, the APF decays quickly around the transition chemical potential,
$\mu=0.35~\mathrm{GeV}$.
Compared with the results without the repulsive-vector type interaction~\cite{Kashiwa2019a},
the reduction of APF is more rapid and stronger.
The complex phase from $\omega$ reads $-2ikG_v\omega\rho_q/T^4$,
which causes cancellation of the Boltzmann weight
at finite quark number density.
With optimization, $\omega$ can take a complex value $\omega \simeq -i \rho_q$,
expected from the mean field results,
and the weight cancellation is weakened.
APF with optimization still decreases to be around 0.6 at $\mu\simeq 0.35~\mathrm{GeV}$,
where both the vector field and the Polyakov loop grow rapidly.
In the middle and right panels of Fig.~\ref{Fig:PNJL1}, 
we show $\sigma/\sigma_0$, $\mathrm{Im}(\omega)/\sigma_0$, $\Phi$ and $\overline{\Phi}$
as functions of $\mu$ at $T=0.1~\mathrm{GeV}$ and $k=64$,
where $\sigma_0$ denotes the vacuum value of $\sigma$.
While APF reduction is seen, these observables seem to be obtained precisely.
The volume factor $k=64$ is large enough and the Monte-Carlo integrals on the optimized path
agree with the mean field results (dashed lines).

\begin{figure}[htbp]
\begin{center}
\includegraphics[width=10cm,bb=35 15 688 260,clip]{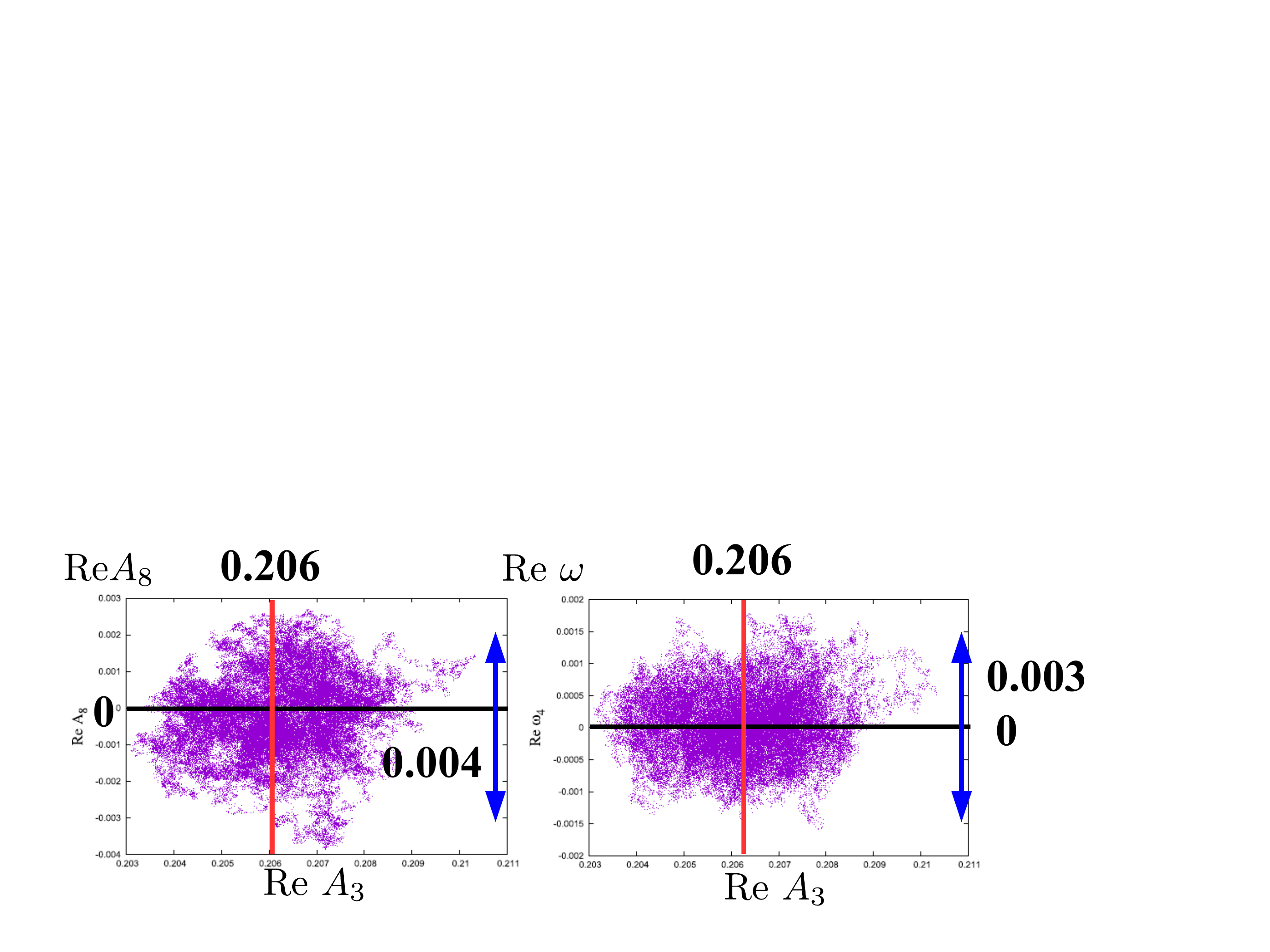}%
\end{center}
\caption{Distribution of the HMC samples
on the $(\mathrm{Re}\,A_3,\mathrm{Re}\,A_8)$ (left) and 
and $(\mathrm{Re}\,A_3,\mathrm{Re}\,\omega)$ (right) planes in the PNJL model.}
\label{Fig:PNJL2}
\end{figure}

In Fig.~\ref{Fig:PNJL2}, we show the distribution of the Monte-Carlo configurations
in the $(\mathrm{Re}\,A_3, \mathrm{Re}\,A_8)$ and $(\mathrm{Re}\,A_3, \mathrm{Re}\,\omega)$ planes.
As the $\mathcal{CK}$ (mean field) ansatz predicts,
$\mathrm{Re}\,A_8 \simeq 0$ ($\mathrm{Re}\,\omega \simeq 0$) 
is confirmed by the Monte-Carlo configurations on the optimized path.
This observation tells us that the statistical weight decreases rapidly 
with finite values of $\mathrm{Re}\,A_8$ or $\mathrm{Re}\,\omega$. 

\section{Summary}
\label{Sec:Summary}

We have discussed the sign problem in the Polyakov loop extended Nambu--Jona-Lasinio (PNJL) model
with repulsive vector-type interaction by using the path optimization
method~\cite{Kashiwa2019b}.
We have confirmed that the model sign problem is weakended
by the path optimization, 
and that the prescriptions adopted in the mean field treatments
are in fact reasonable.

We have also derived the Euler-Lagrange equation,
which should be satisfied by the optimal path.
The variationally optimized path is found to agree
with the solution of the Euler-Lagrange equation
in the region with large statistical weight
in the one-site Hubbard model.
The subset resummation is also discussed as a prescription
to enhance APF over the upper bound on the single path.

\bigskip

This work is supported in part
by the Grants-in-Aid for Scientific Research from JSPS (Nos. 
18J21251, 
18K03618, 
19H01898, 
and
19H05151), 
and by the Yukawa International Program for Quark-hadron Sciences (YIPQS).

\end{document}